\documentclass[reprint, aps, pra, floatfix]{revtex4-2}
\usepackage[dvips]{graphicx}
\usepackage[english]{babel}
\usepackage{amsmath}
\usepackage{amssymb}
\usepackage{mathrsfs}
\usepackage{bm}
\usepackage{subfigure}
\usepackage[colorlinks, citecolor = black, linkcolor = black, urlcolor = blue]{hyperref}

\newcommand{\bk}{\mathbf{k}}

\newcommand{\nn}{\nonumber}
\newcommand{\x}{\hat{\mathcal{X}}}
\newcommand{\hk}{\mathcal{H}_\mathbf{k}}
\newcommand{\h}{\mathcal{H}}
\newcommand{\cua}{\mathcal{U}}
\newcommand{\cva}{\mathcal{V}}
\newcommand{\cu}{\mathcal{U}(t)}
\newcommand{\cv}{\mathcal{V}(t)}

\newcommand{\va}{|\mathbf{0}\rangle}

\newcommand{\hp}{\hat{\Psi}}
\newcommand{\hu}{\hat{U}}

\begin{document}
\title{Quantum Echo in Two-Component Bose-Einstein Condensates}
\author{Chang-Yan Wang}
\email{changyanwang@tsinghua.edu.cn}
\affiliation{Institute for Advanced Study, Tsinghua University, Beijing 100084, China}
\begin{abstract}
  The development of ultracold atom technology has enabled the precise investigations on quantum dynamics of quantum gases. Recently, inspired by experimental advancement, the $SU(1,1)$ echo, akin to the well-known $SU(2)$ spin echo, has been proposed for single-component Bose-Einstein condensate (BEC). In this paper, we investigate the possibility of quantum echo in the more intricate two-component BEC by fully exploiting its underlying symmetry, which is the Lie group $Sp(4,R)$. We demonstrate that quantum echo can occur for the two-component BEC by applying a driving protocol consisting of two steps in each period. The first step can be any Bogoliubov Hamiltonian, while the second step is a Hamiltonian with interactions turned off, which plays a similar role as the $\pi$-pulse in spin echo. We confirm our theoretical results with numerical calculations for different examples of two-component BEC. We further consider the effect of interactions between the excited boson modes on the quantum echo process and discuss the possible experiment implementation of this quantum echo.
\end{abstract}
\maketitle

\section{Introduction}
In the past decades, ultracold-atomic systems have emerged as a powerful tool for the precise exploration of dynamics within quantum systems, as they offer high controllability and tunability \cite{bloch_ultracold_2005, bloch_manybody_2008, chin_feshbach_2010}. Recently, The Chicago group's experiments on Bose-Einstein condensates (BECs) have revealed exotic phenomena like Bose fireworks and quantum revival of BEC \cite{clark_collective_2017, feng_correlations_2019, fu_density_2018, clark_quantum_2015, clark_universal_2016, feng_coherent_2018, fu_jet_2020, zhang_pattern_2020, hu_quantum_2019}. The latter one inspired the study of quantum echo, i.e. the revival of the quantum state of a system after applying a specific driving protocol in single-component BEC using the $SU(1,1)$ group \cite{chen_manybody_2020, lv_su_2020, lyu_geometrizing_2020, zhang_quantum_2022, cheng_manybody_2021}, which resembles the $SU(2)$ spin echo \cite{hahn_spin_1950}.

On the other hand, BEC can also form in more complex systems consisting of two or more species (internal states) of bosonic atoms, such as the two-component and spinor BEC \cite{ho_spinor_1998, kawaguchi_spinor_2012, ohmi_boseeinstein_1998}. The two-component BEC has been realized in various ultracold atom systems \cite{myatt_production_1997,hall_dynamics_1998,hall_measurements_1998, modugno_two_2002, papp_tunable_2008,thalhammer_double_2008, mccarron_dualspecies_2011, lercher_production_2011, pasquiou_quantum_2013, wacker_tunable_2015, wang_double_2015, trautmann_dipolar_2018}, and received extensive investigations \cite{ho_binary_1996, esry_hartreefock_1997, alon_multiconfigurational_2008, ao_binary_1998, cirac_quantum_1998, mertes_nonequilibrium_2007, pu_properties_1998, timmermans_phase_1998, micheli_manyparticle_2003, tojo_controlling_2010, trippenbach_structure_2000, wang_spinorbit_2010}. Given the tunability of the interaction between different species of atoms through Feshbach resonances \cite{thalhammer_double_2008}, one may wonder whether such a quantum echo can occur in this more complicated two-component BEC system.

In this paper, we address this problem by using the general formalism presented in Ref.\cite{wang_quantum_2022}, which employs the so-called real symplectic group $Sp(4,R)$ to deal with the quantum dynamics of the two-component BEC. The $Sp(4,R)$ group, a non-compact Lie group, preserves the canonical commutation relations of boson operators \cite{perelomov_generalized_1986} and finds applications across various areas of physics, such as quantum information \cite{weedbrook_gaussian_2012, simon_pereshorodecki_2000, simon_gaussian_1988}, high energy physics \cite{colas_fourmode_2022, alhassid_group_1983, deenen_boson_1985, enayati_antide_2023}, and cold atoms \cite{penna_twospecies_2017, richaud_quantum_2017, charalambous_control_2020, wang_quantum_2022}.  (See Appendix \ref{app:sp} for a brief introduction to this group.) In Ref.\cite{wang_quantum_2022}, by mapping the time evolution operator of BEC to an $Sp(4,R)$ matrix, the quantum dynamics under arbitrary Bogoliubov Hamiltonian can be calculated.

Utilizing this formalism, we show that through a two-step periodic driving protocol, the fully condensed state or any $Sp(4,R)$ coherent state of a two-component BEC can revert to its original form after two driving periods. The first-step Hamiltonian $\hat{H}_1$ can be any Bogoliubov Hamiltonian, while in the second step $\hat{H}_2$, the interactions are turned off. We provide a method to calculate $\hat{H}_2$ for a given $\hat{H}_1$. This is demonstrated through two examples: a time-independent $\hat{H}_1$ and a time-dependent $\hat{H}_1$, with numerical results confirming our theory. We also examine the quantum echo breaking effect due to interactions between excited boson modes and analyze how this effect varies with the kinetic to interaction strength ratio in the first-step Hamiltonian. The paper concludes with a discussion on potential implementations of these quantum echoes in cold atom experiments.

Our work highlights the exploitation of the underlying symmetry of two-component BEC. Since the similar symmetry properties shared by $N$-component BEC, our work provide a unified perspective on the quantum echoes of one and two-component BEC, and can be potentially generalized to the spinor BEC case. Furthermore, given the presence of real symplectic groups in various areas of physical study \cite{simon_pereshorodecki_2000, simon_gaussian_1988, colas_fourmode_2022, alhassid_group_1983, deenen_boson_1985, enayati_antide_2023, penna_twospecies_2017, richaud_quantum_2017, charalambous_control_2020}, especially the bosonic Gaussian states \cite{weedbrook_gaussian_2012}, our work can shed light on the revival of quantum states in these research areas.

This paper is organized as follows: in Sec. \ref{formalism}, we review the formalism using $Sp(4,R)$ group to deal with the quantum dynamics of two-component BEC; in Sec. \ref{echo}, we prove that a two-step driving protocol can make the quantum echo occur; in Sec. \ref{numerical}, we apply this driving protocol to two different examples, and present the numerical results to confirm our proof; in Sec. \ref{interaction}, we consider the interaction effect; in Sec. \ref{experiment}, we propose a possible experimental implementation of this echo; we conclude in Sec. \ref{conclusion}.

\section{The quantum dynamics and $Sp(4,R)$ symmetry}\label{formalism}
We consider the physical process that a two-component Bose gas is prepared in a fully condensed state, i.e. all particles condensed in the zero momentum modes, then the system evolves under the Bogoliubov Hamiltonian. The general form of this quench Bogoliubov Hamiltonian can be written as $\hat{H}_\mathrm{Bg}(t) = \sum_{\bk\neq \mathbf{0}} \hat{H}_\bk(t) + \mathrm{const.}$, where
\begin{eqnarray} \label{ham}
  \hat{H}_\bk(t) = \hp_\bk^\dagger \hk(t) \hp_\bk, \ \hk(t) = \left( \begin{array}{cc} \xi_\bk(t) & \eta(t) \\ \eta^*(t) & \xi_\bk^*(t) \end{array} \right),
\end{eqnarray}
with $\hp_\bk = (a_{1,\bk}, a_{2,\bk}, a_{1,-\bk}^\dagger, a_{2,-\bk}^\dagger)^T$ \cite{pitaevskii_boseeinstein_2016, wang_quantum_2022}. Here $\xi_\bk(t)$ is a $2\times 2$ Hermitian matrix given by $\xi_{\bk, ij}(t) = \frac{1}{2}[\varepsilon_{\bk,ij}(t) + g_{ij}(t) \psi_i \psi_j^*]$, and $\eta(t)$ is a complex symmetric matrix given by $\eta_{ij}(t) = \frac{1}{2} g_{ij}(t) \psi_i\psi_j$, and $i,j = 1,2$ label the two species of bosons. The diagonal elements of $\varepsilon_\bk(t)$ are the kinetic terms, whereas the off-diagonal elements are the spin-orbit coupling terms. $g_{ij}(t)$ represent the intra- and inter-species interaction strengths of these two species of bosons. $\psi_i = \sqrt{N_i/V}e^{i\theta}$ are the condensate wavefunctions of the zero momentum modes. And for simplicity, we set $\hbar = 1$.

On the other hand, the two species of the boson operators naturally give a representation of the Lie algebra of the real symplectic group $Sp(4,R)$ \cite{perelomov_generalized_1986,hasebe_sp_2020}. The generators of this Lie algebra can be defined as 
\begin{eqnarray}\label{liealg}
&&\x_{ij} = a_{i,\bk} a_{j,-\bk} + a_{j,\bk} a_{i,-\bk}, \nn\\ 
&&\x^{ij} = a_{i, \bk}^\dagger a_{j,-\bk}^\dagger + a_{j, \bk}^\dagger a_{i,-\bk}^\dagger, \nn\\
&&\x_l^k = a_{k, \bk}^\dagger a_{l,\bk} + a_{l,-\bk} a_{k,-\bk}^\dagger,
\end{eqnarray}
where $i,j,k,l = \{1,2\}$, and ten of them are independent. The Casimir operator of this Lie algebra is given by \cite{hasebe_sp_2020}
\begin{eqnarray}
  \mathcal{C} = \x^{ij}\x_{ij} + \x_{ij}\x^{ij} - 2\x_j^i\x_i^j.
\end{eqnarray}
To show that this is a conserved quantity, we can rewrite it as \cite{hasebe_sp_2020}
\begin{eqnarray}
  \mathcal{C} = -2(\Delta N_\bk +2)(\Delta N_\bk - 2),
\end{eqnarray}
where $\Delta N_\bk = \sum_{i=1}^2 (a_{i,\bk}^\dag a_{i,\bk} - a_{i,-\bk}^\dag a_{i,-\bk})$ counts the difference in the number of bosons between the momenta $+\bk$ and $-\bk$. According to the definition of $\mathfrak{sp}(4,R)$ generators Eq.(\ref{liealg}), none of these generators change this number difference. Thus, the Casimir operator is conserved.

These operators have a one-to-one correspondence with the matrix representation of the $\mathfrak{sp}(4,R)$ Lie algebra. This correspondence can be revealed by rewriting the operators defined in Eq.(\ref{liealg}) as
\begin{eqnarray}
  \x_{ij}=\hp_\bk^\dagger \kappa Y_{ij} \hp_\bk,\ \x^{ij}=\hp_\bk^\dagger \kappa Y^{ij} \hp_\bk,\ \x_l^k=\hp_\bk^\dagger \kappa Y_l^k \hp_\bk,
\end{eqnarray}
where $\kappa = \mathrm{diag}(1,1,-1,-1)$ \cite{hasebe_sp_2020,wang_quantum_2022}. Here, these $4\times 4$ matrices $\{Y_{ij}, Y^{ij}, Y_l^k\}$ are the matrix representation of the Lie algebra $\mathfrak{sp}(4,R)$, and their explicit form is given in Appendix \ref{app1}. Then, we can write the Bogoliubov Hamiltonian $\hat{H}_\bk$ in terms of these generators as
\begin{eqnarray} \label{hkt}
  \hat{H}_\bk(t) = \xi_{\bk,ij}(t) \x_j^i + \frac{1}{2} \eta_{ij}(t)\x_{ij} + \frac{1}{2} \eta_{ij}^*(t) \x^{ij},
\end{eqnarray}
where we have adopted the Einstein summation convention, i.e. the repeated indices $i$ and $j$ are summed over. In the following, we always adopt this convention implicitly. Hence, the time evolution operator 
\begin{eqnarray}
  \hu_\bk(t) = \mathcal{T} e^{-i\int_0^t \hat{H}_\bk(t') dt'}
\end{eqnarray}
gives a representation of the real symplectic group $Sp(4,R)$, i.e. $\hat{U}_\bk(t)$ has a one-to-one correspondence to a matrix in the group $Sp(4,R)$, which is given by 
\begin{eqnarray}
  U_\bk(t) = \mathcal{T} e^{-i\int_0^t \kappa \hk(t')dt'}.
\end{eqnarray} 
Here, $\mathcal{T}$ is the time ordering operator. 
This can be seen by noticing that the matrix corresponding to the operator $\hat{H}_\bk(t)$ is
\begin{eqnarray}
  \kappa \h_\bk(t) = \xi_{\bk,ij}(t) Y_j^i + \frac{1}{2} \eta_{ij}(t) Y_{ij} + \frac{1}{2} \eta_{ij}^*(t) Y^{ij}. \nn
\end{eqnarray}
Since the matrix $U_\bk(t)$ is a real symplectic matrix, it has the form \cite{perelomov_generalized_1986}
\begin{eqnarray}\label{ukt}
  U_\bk(t) = \left( \begin{array}{cc} \cu  & \cv \\ \cv^* & \cu^* \end{array} \right),
\end{eqnarray}
where $\cu, \cv$ are $2 \times 2$ matrices satisfying
\begin{eqnarray}\label{constraints}
  \cu \cu^\dagger - \cv\cv^\dagger = I,\ \cu \cv^T = \cv\cu ^T. 
\end{eqnarray}

In order to calculate the time evolution of the fully condensed state, one can decompose $\hat{U}_\bk(t)$ as 
\begin{eqnarray} \label{sp_decom}
  \hat{U}_\bk(t) = e^{-\frac{1}{2} Z_{ij}(t) \x^{ij}} e^{\zeta_{kl}(t) \x_l^k} e^{-\frac{1}{2}\nu_{ij}(t)\x_{ij}},
\end{eqnarray}
which is called normal order decomposition \cite{perelomov_generalized_1986,wang_quantum_2022}. With this decomposition, the fully condensed state $\va$ evolves as 
\begin{eqnarray}
  \hat{U}_\bk(t)\va = \mathcal{N_\bk} e^{-\frac{1}{2} Z_{ij}(t)\x^{ij}} \va \equiv |Z(t)\rangle,
\end{eqnarray}
where $|Z(t)\rangle$ is an $Sp(4,R)$ coherent state and $\mathcal{N_\bk}$ is normalization factor. At any time $t$ of the evolution, $Z(t)$ is \cite{rowe_vector_1985, wang_quantum_2022}
\begin{eqnarray}
  Z(t) &=& \cv[\cu^*]^{-1},
\end{eqnarray}
where $\cu, \cv$ is defined in Eq.(\ref{ukt}). By using the Eq.(\ref{constraints}), it can be shown that at any time t, $Z(t)$ is a $2\times 2$ symmetric complex matrix, and $I - Z(t)^\dag Z(t)$ is positive definite, where $I$ is $2\times 2$ identity matrix.  Thus, the matrix $Z(t)$ can be parameterized by three complex numbers $Z_{11}(t), Z_{12}(t), Z_{22}(t)$, or by six real parameters. Mathematically, $Z(t)$ lies in a six-dimensional manifold called Cartan classical domain, and is isomorphic to the quotient space $Sp(4,R)/U(2)$ \cite{coquereaux_conformal_1990, perelomov_generalized_1986}, where $U(2)$ is the unitary group of degree 2. Thus, $Z(t)$ gives a trajectory in this six-dimensional manifold. And if the initial state is arbitrary coherent state $|Z_0\rangle$, the time evolution of this coherent is also a coherent state, i.e. $\hat{U}_\bk(t)|Z_0\rangle = |Z'(t)\rangle$, and $Z'(t)$ is given by \cite{wang_quantum_2022, rowe_vector_1985}
\begin{eqnarray} \label{zpt}
  Z'(t) = [\cu Z_0 + \cv][\cv^* Z_0 + \cu^*]^{-1}.
\end{eqnarray}
With this general formalism dealing with the time evolution of two-component BEC in hand, we are ready to study the quantum echo of the two-component BEC system.

\section{The quantum echo}\label{echo}
\begin{figure}\label{protocol}
  \includegraphics[width = .35\textwidth]{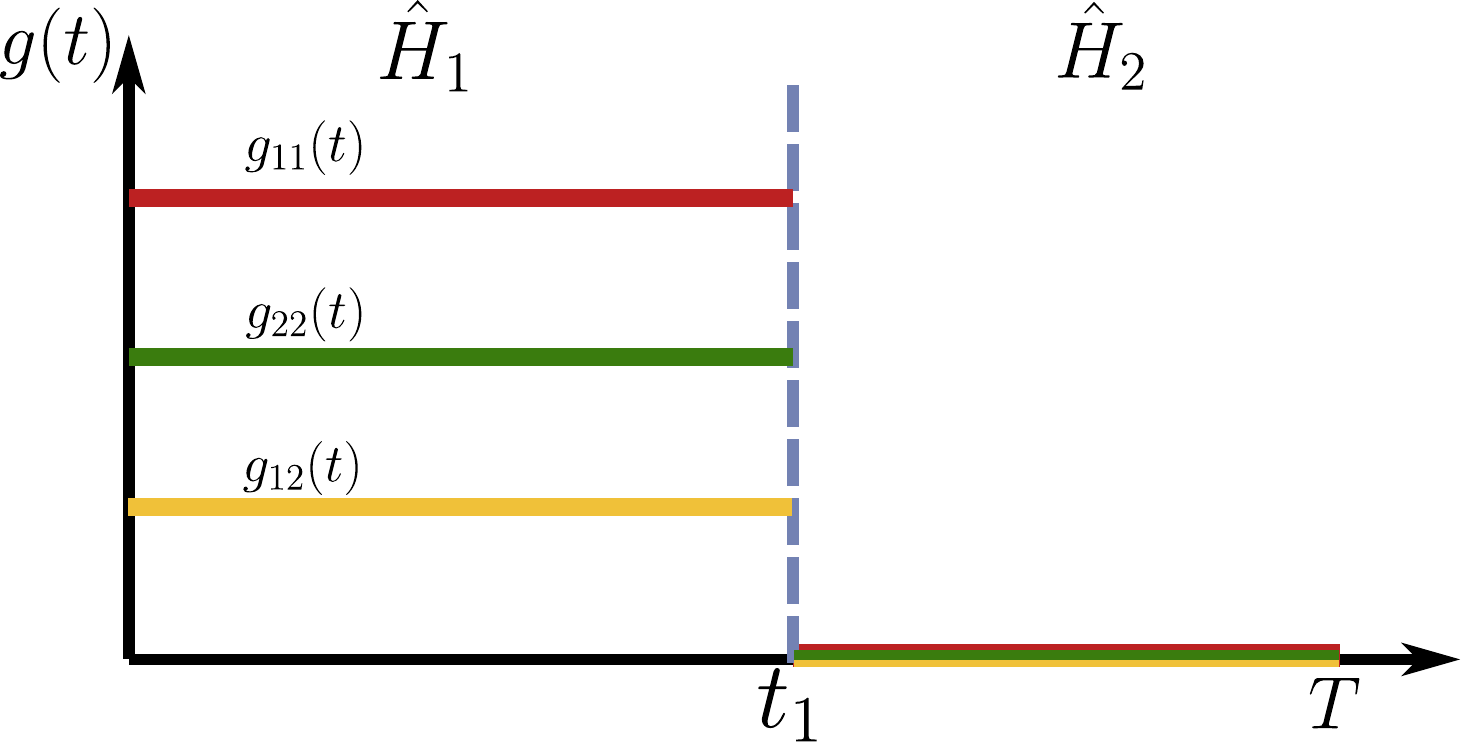}
  \caption{Schematics for the driving protocol. In the first step of a period, the Hamiltonian can be any Bogoliubov Hamiltonian. However, in the second step of a period, the all the interaction strengths should be turned off.}
  \label{}
\end{figure}

In this section, we will study the problem of quantum echo for boson modes in generic $\pm\bk \neq \mathbf{0}$ momenta by considering a periodic driving protocol similar to Ref.\cite{chen_manybody_2020,lyu_geometrizing_2020}. In this driving protocol, each period consists of two steps: in the first step, the Hamiltonian $\hat{H}_1$ is a generic two-component Bogoliubov Hamiltonian defined in Eq.(\ref{hkt}); in the second step, the Hamiltonian $\hat{H}_2$ only includes these $\x_j^i$ operators, i.e. turning off all the intra- and inter-species interaction strengths $g_{ij}$, as shown in Figure \ref{protocol}. We will show how to obtain the explicit form of the second-step Hamiltonian $\hat{H}_2$ that make the initial state, either $\va$ or a generic coherent state $|Z_0\rangle$, reverse to itself after two period of driving for any given Bogoliubov Hamiltonian $\hat{H}_1$.

In the $m$th period, when $(m-1)T < 0 \leq (m-1)T+t_1$, the Hamiltonian is
\begin{eqnarray}
  \hat{H}_1 &=& \xi_{ij}(t) \x_j^i + \frac{1}{2} \eta_{ij}(t)\x_{ij} + \frac{1}{2} \eta_{ij}^*(t) \x^{ij}, 
\end{eqnarray}
and when $(m-1)T + t_1 < t \leq mT$
\begin{eqnarray}
  \hat{H}_2 &=& \frac{1}{2} \varepsilon_{ij} \x_j^i,
\end{eqnarray}
where $T = t_1+t_2$ is the duration of each period. Here we have omitted the subscript $\bk$ for simplicity. Thus, for such a quantum echo to occur, we require that the time evolution operators $\hat{U}_1(t_1)$ and $\hat{U}_2(t_2)$ of these two steps acting on any coherent state $|Z_0\rangle$ gives
\begin{eqnarray}
  (\hat{U}_2(t_2) \hat{U}_1(t_1))^2 |Z_0\rangle = |Z_0\rangle.
\end{eqnarray}
According to Eq.(\ref{zpt}), this condition is equivalent to
\begin{eqnarray}\label{usqr}
  U(T)^2 = (U_2(t_2)U_1(t_1))^2 = \pm 1,
\end{eqnarray} 
where $U_i(t_i), i = 1,2$ are the time evolution matrix corresponding to $\hat{U_i}(t_i)$. Next we will focus the case $U(T)^2 = -1$, since the result for $U(T)^2 = 1$ case can be obtained by redefined $U(T)^2$ from $U(T)^2 = -1$ case as $U(T)$.

Since $U(T)$ is also a real symplectic matrix i.e. having the form of Eq.(\ref{ukt}), substituting its elements $\cua, \cva$ into Eq.(\ref{usqr}) leads to
\begin{eqnarray}\label{condi}
  \cua^2 + \cva\cva^* &=& \cua^{*2} + \cva^*\cva = -1, \nn\\
  \cua\cva + \cva \cua^* &=& 0.
\end{eqnarray}
By using the property of real symplectic matrix $ \cua \cua ^\dagger - \cva\cva^\dagger = I,\ \cua \cva^T = \cva\cua ^T$, the Eqs.(\ref{condi}) result in
\begin{eqnarray}\label{uudag}
  \cua = -\cua^\dag.
\end{eqnarray}
This is the condition that $U(T)$ should satisfy for the quantum echo to occur.

Next, we will construct an explicit form of $U(T)$. We recall that for the $SU(1,1)$ case, an $SU(1,1)$ matrix satisfying $U^2 = -1$ can be given by
\begin{eqnarray}
  (e^{i \pi \frac{\sigma_z}{2}}e^{i (\alpha_+ \sigma^+ - \alpha_- \sigma^-)})^2 = -1,
\end{eqnarray}
where $\{\sigma_z/2,\sigma^+,-\sigma^-\}$ are the generators of $SU(1,1)$ group, with $\sigma_{x,y,x}$ the Pauli matrices and $\sigma^\pm = 1/2(\sigma_x \pm i\sigma_y)$, and $\alpha_\pm$ are arbitrary complex numbers. We notice the resemblance in form between $\{\sigma_z,\sigma^+,-\sigma^-\}$ and $\{ Y_l^k, Y_{ij}, Y^{ij}\}$. Inspired by this resemblance, we expect that for the $Sp(4,R)$ case, the matrix $U(T)$ satisfying $U(T)^2 = -1$ is given by 
\begin{eqnarray}\label{ut}
  U(T) &=& e^{i\frac{\pi}{2} (Y_1^{1} + Y_2^2)} e^{\frac{i}{2}(\eta_{ij} Y^{ij} + \eta_{ij}^* Y_{ij})}
\end{eqnarray}
where $\eta$ is a $2\times 2$ symmetric complex matrix. Since $e^{i\frac{\pi}{2} (Y_1^{1} + Y_2^2)} = \mathrm{diag}(i,i,-i,-i)$, substituting the above equation to Eq.(\ref{uudag}) leads to that the sub-matrix $\cua_0$ of $U_0 = e^{\frac{i}{2}(\eta_{ij} Y^{ij} + \eta_{ij}^* Y_{ij})}$ should be Hermitian, i.e. $\cua_0 = \cua_0^\dag$. In Appendix \ref{app1}, we give the explicit expression of $\cua_0$ and also prove that $\cua_0$ is Hermitian. As a result, we have proven that the form of $U(T)$ given in Eq.(\ref{ut}) actually satisfy $U(T)^2 = -1$.

\begin{figure}
  \includegraphics[width = .45\textwidth]{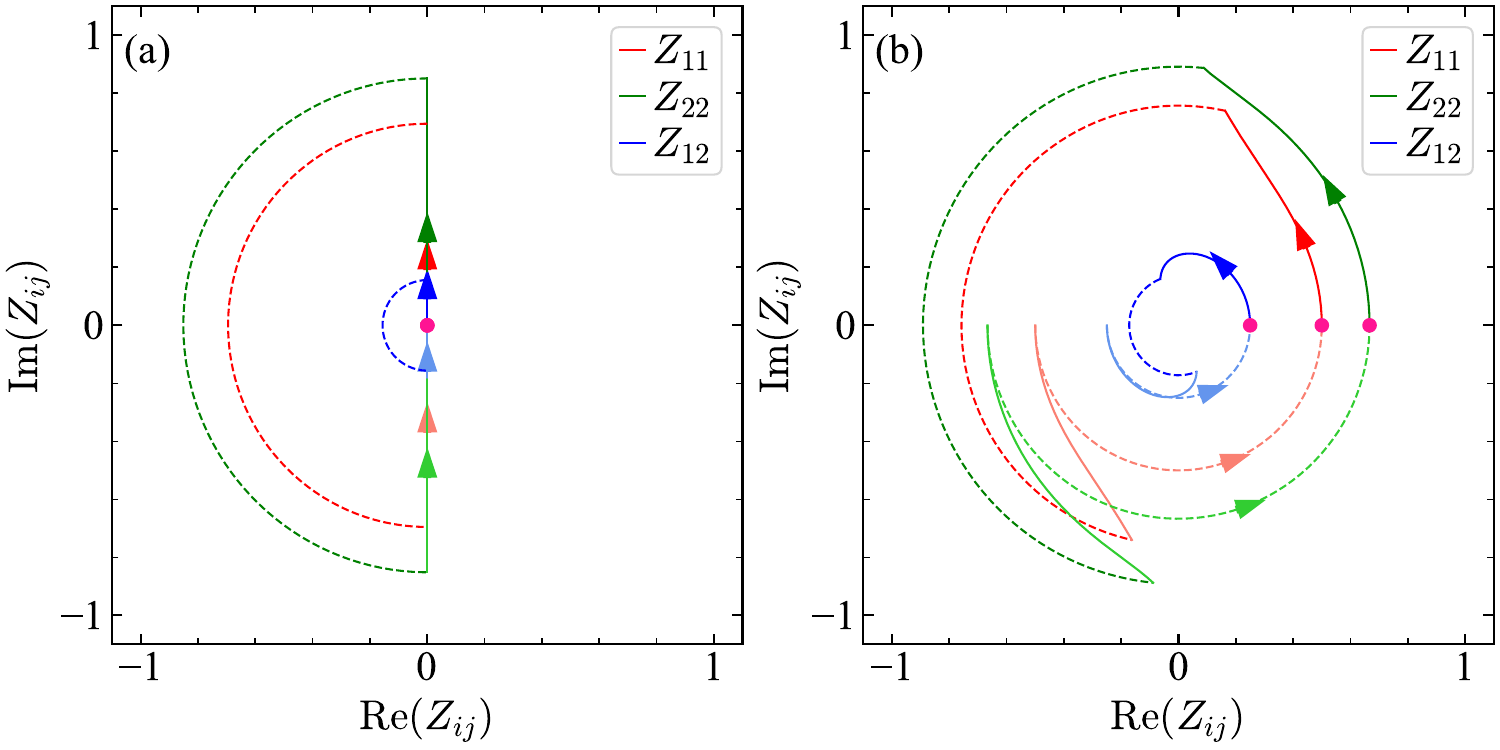}
  \caption{Trajectories of the elements of $Z(t)$ when $\eta_{11} = 1, \eta_{22} = 3/2, \eta_{12} = 1/2$ for (a) the initial state is $\va$, and (b) the initial state is a coherent state $|Z_0\rangle$ with $Z_0 = \left(\begin{array}{cc} 1/2 & 1/4 \\ 1/4 & 2/3 \end{array} \right)$. The evolution includes two periods. The lines (both solid and dashed) with darker colors represent the first period, and the line with lighter colors represent the second period. The solid lines represent the first steps and the dashed lines represent the second steps. }
  \label{proto}
\end{figure}

\begin{figure*}
  \includegraphics[width = \textwidth]{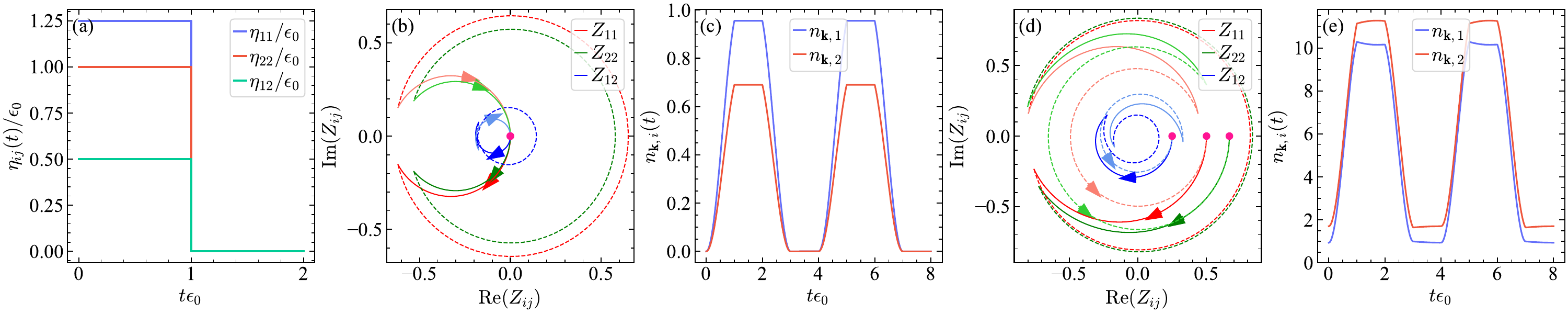}
  \caption{Quantum echo when the interaction strengths are all constant in the first step of each period. (a) The interaction terms $\eta_{ij}(t)$ in one period. (b) The trajectory of the $Z(t)$ matrix in the parameter space when initial state is $\va$. The purple dot represents the initial state, the arrows represent the direction of the trajectories as time evolves. The solid lines represent the first step and the dashed lines represent the second step. The lines with deep colors are the first period and light-colored ones represent the second period. (c) The time evolution of the particle number for each species of bosons when initial state is a vacuum. (d), (e) the same Hamiltonian as (c) and (d) but the initial is a generic coherent state $|Z_0\rangle$ which is the same as the one in Figure \ref{proto}.} 
  \label{reviv}
\end{figure*}

With the general form of $U(T)$, we can find the $\hat{H}_2$ that make the quantum echo occur for arbitrary $\hat{H}_1$. Since $\hat{H}_1$ is of the general form of Eq.(\ref{hkt}), its corresponding time evolution matrix is also of the general form of real symplectic matrix Eq.(\ref{ukt}), i.e.
\begin{eqnarray}
  U_1(t_1) = \left(\begin{array}{cc} \cua_1(t_1) & \cva_1(t_1) \\ \cva_1(t_1)^* & \cua_1(t_1)^* \end{array}\right).
\end{eqnarray} 
However, $\hat{H}_2$ only has $\x_j^i$ terms, thus, its corresponding time evolution matrix is block diagonal, i.e.
\begin{eqnarray}
  U_2(t_2) = \left(\begin{array}{cc} \cua_2(t_2) & 0 \\ 0 & \cua_2(t_2)^* \end{array}\right).
\end{eqnarray}
And the constraints on the real symplectic matrix Eq.(\ref{constraints}) also results in that $\cua_2(t_2)$ is unitary. Then, by substituting the elements of $U_1(t_1)$ and $U_2(t_2)$ into $U(T) = U_2(t_2)U_1(t_1)$, we have
\begin{eqnarray}\label{polar}
  -i\cua_1(t_1) = \cua_2(t_2)^{-1}\cua_0.
\end{eqnarray}
Since we have shown that $\cua_2(t_2)^{-1}$ is unitary and $\cua_0$ is Hermitian, the right-hand side of the above equation is just the polar decomposition of $-i\cua_1(t_1)$. The polar decomposition of a complex matrix is the factorization of this matrix into a product of unitary matrix and a Hermitian matrix. And the polar decomposition of a square matrix always exist. Hence, by doing the polar decomposition we can find $\cua_2(t_2)$. And the explicit form of $\h_2$ is given by 
\begin{eqnarray}\label{h2}
  \h_2 = \kappa\ln(\cua_2(t_2))/(-it_2),
\end{eqnarray}
where $\kappa = \mathrm{diag}(1,1,-1,-1)$.

To have a better understanding of why we choose $U(T)$ to have the form of Eq.(\ref{ut}), we can let $\hat{H}_1 = -\frac{1}{2}(\eta_{ij}\x^{ij} + \eta_{ij}^*\x_{ij})$, $\hat{H}_2 = -\frac{\pi}{2}(\x_1^1 + \x_2^2)$, and the times $t_1 = t_2 = 1$. Figure (\ref{proto}) shows the trajectory of the elements of the matrix $Z_t$ in the complex plane for the cases when the initial state is $\va$ and a generic coherent state $|Z_0\rangle$ respectively. It can be seen that in both cases, all the trajectories perfectly return to their initial points, which confirms above proof. From this figure, we can see that for both cases, the second step $\hat{H}_2$ amounts to a $\pi$ rotation for all the elements of the $Z_t$ matrix. It reverses the direction of the trajectories and make the trajectories return to its initial points. Thus, we can see that the second-step Hamiltonian plays a similar role to the $\pi$ pulse in the $SU(2)$ spin echo. Meanwhile, when the initial state is $\va$, the second step in the second period is not necessary for quantum echo, and applying $\hat{H}_1, \hat{H}_2$ and $\hat{H}_1$ in consequence suffices to make the state return to its initial state. However, if the initial state is a generic coherent state, the full driven protocol is necessary for the quantum echo to happen.

\section{Numerical results}\label{numerical}
In previous section, we have provided a protocol that can bring any coherent state back to itself after two periods of driving for any Bogoliubov Hamiltonian $\hat{H}_1$. Briefly, in this protocol, each period consists two steps: the first step can be any Bogoliubov Hamiltonian $\hat{H}_1$, and the second step is a quadratic Hamiltonian $\hat{H}_2$ with all interaction strengths turned off. The Hamiltonian in the second step can be found by doing a polar decomposition for the $2\times 2$ sub-matrix of the time evolution matrix $U_1(t_1)$ as shown in Eq.(\ref{polar}) and Eq.(\ref{h2}). 

\begin{figure*}
  \includegraphics[width = \textwidth]{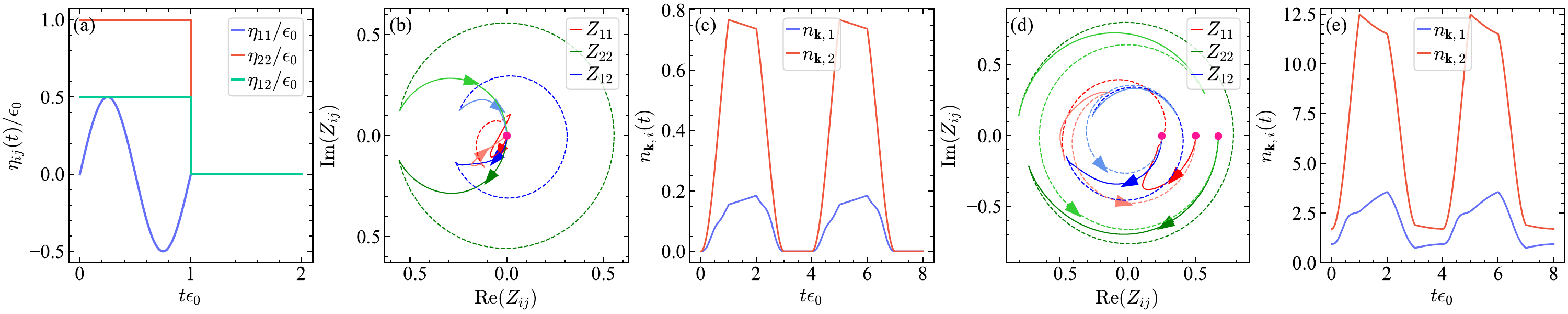}
  \caption{Quantum echo when one interaction strength is a sine wave. (a) The interactions $g_{ij}(t)$ in a period. (b) The trajectory of the $Z(t)$ matrix in the parameter space when initial state is the vacuum. The convention in this figure is the same as Figure \ref{reviv}. (c) The time evolution of the particle number of each species of bosons when initial state is a vacuum. (d), (e) the same Hamiltonian as (c) and (d) but the initial is a generic coherent state $|Z_0\rangle$, which is the one we used in Figure \ref{proto}.}
  \label{reviv_sine}
\end{figure*}

In this section, we will apply this protocol to some examples and present the numerical results. The first example is the case when the interaction strengths all keep constant in the first step of a period as shown in Figure \ref{reviv}. For simplicity, in the first step of each driving period, we set the masses of the two species of bosons equal and set the spin-orbit coupling as 0, more specifically we set $\varepsilon_\bk = \mathrm{diag}(\epsilon_0,\epsilon_0)$, where the matrix $\varepsilon_\bk$ is defined in the Eq.(\ref{ham}). Here, $\epsilon_0 = \bk^2/2m_1$ is the kinetic energy of the first boson mode, and serves as an energy unit. Although we set the boson masses equal for simplicity, our approach also applies to the general case with unequal boson masses, as demonstrated analytically in the previous section. And the interaction terms $\eta_{ij}$ are chosen as $\eta_{11} = 1.25 \epsilon_0,\ \eta_{22} = \epsilon_0,\ \eta_{12} = 0.5 \epsilon_0$ as shown in Figure \ref{reviv}(a). And we choose the duration of each step equal i.e. $t_1 = t_2 = 1/\epsilon_0$. According to Eq.(\ref{polar}) and Eq.(\ref{h2}), the Hamiltonian in the second step is
\begin{eqnarray}
  \h_2 = \left(\begin{array}{cc} \varepsilon & 0 \\ 0 & \varepsilon^* \end{array} \right), \ \varepsilon \simeq -\left(\begin{array}{cc} 2.85 & 0.21 \\ 0.21 & 2.75 \end{array} \right) \epsilon_0.
\end{eqnarray}
Here, we can see that with the interaction strengths turned off, only the kinetic terms and the spin-orbit coupling terms left in $\h_2$. However, the diagonal terms in $\varepsilon$ are no longer equal, which means different chemical potentials for these two species of bosons are required. Meanwhile, the off-diagonal terms are also no longer $0$, requiring non-zero spin-orbit coupling. 

The numerical results for this case are shown in Figure \ref{reviv}. Figure \ref{reviv}(a) shows the values of the interaction strengths. Figure \ref{reviv}(b) shows the evolution of the three independent complex elements of the symmetric matrix $Z(t)$ in the complex plane when the initial state is the fully condensed state $\va$. In Figure \ref{reviv}(c), we show the evolution of the particle number of each component of boson, which can be calculated using the method presented in Ref.\cite{wang_quantum_2022}. Figure \ref{reviv}(d) and (e) show the trajectories and particle numbers when the initial state is a generic non-vacuum coherent state. 

The second example is the case when one of the interaction strengths is a sine wave in the first step of each period as shown in Figure \ref{reviv_sine}. Here, we still choose the $\varepsilon_\bk$ matrix in the first-step Hamiltonian as diagonal, i.e. $\varepsilon_\bk = \mathrm{diag}(\epsilon_0,\epsilon_0)$, and set the duration of each step equal i.e. $t_1 = t_2 = {red}1/\epsilon_0$. For the interaction terms, $\eta_{11} = \epsilon_0\sin(2\pi \epsilon_0 t)/2,\ \eta_{22} = \epsilon_0$ and $\eta_{12} = 0.5 \epsilon_0$, as shown in Figure \ref{reviv_sine}. Using the same procedure, we can calculate the second-step Hamiltonian $\h_2 = \mathrm{diag}(\varepsilon, \varepsilon^*)$, where
\begin{eqnarray}
 \varepsilon \simeq -\left(\begin{array}{cc} 2.02 & 0.36-0.009i \\ 0.36+0.009i & 2.73 \end{array} \right) \epsilon_0.
\end{eqnarray}
In Figure \ref{reviv_sine}, we show the evolution trajectories in the parameter space and the evolution of particle numbers when the initial state is a fully condensed state and a generic coherent state.

From the numerical results present in Figure \ref{reviv} and \ref{reviv_sine}, we can see that our formalism can clearly display how the initial state evolves in the whole driving process. And it is clearly shown that for these two examples, the driving protocol can bring both the fully condensed state and a generic coherent state back to themselves after two periods. Meanwhile, in these examples, the trajectories of the case with a generic coherent state as the initial state are more complicated than the case with a fully condensed state as initial state. Furthermore, we also show the time evolution of the particle numbers of the two species of bosons. The particle numbers increase rapidly in the first step of the first period, and then in the second step they cease to grow. In the first step of the second period, the particle numbers start to decrease. Finally, they return to their initial values. Thus, numerical results clearly demonstrate how the quantum echo happens in the two-component BEC systems.

\section{Interaction effect}\label{interaction}
In this section, we will consider the interaction effect in the quantum echo process. Similar to the one component BEC \cite{lyu_geometrizing_2020}, the interaction between these boson modes in two-component BEC can be given as $\hat{V} = \sum_{i,j}(4\hat{n}_{i,\bk}\hat{n}_{j,-\bk} + \hat{n}_{i,\bk}\hat{n}_{j,\bk} + \hat{n}_{i,-\bk}\hat{n}_{j,-\bk})$. When there are a large amount of boson modes excited in the $\pm \bk$ momenta, the interaction between these excitations cannot be ignored. As a result, the quantum echo process will be broken. By calculating the particle numbers in the end of $2m$th period, we want to reveal the interplay of the interactions between boson modes and the parameters in the original first-step Hamiltonian $\hat{H}_1$ in this quantum echo breaking process. 

For simplicity, we consider the following two-step driving
\begin{eqnarray}
  \hat{H}_1' &=& \hat{H}_1 + \tilde{g} \hat{V}, \nn\\
  \hat{H}_2 &=& \frac{1}{2} \varepsilon_{ij} \x_j^i,
\end{eqnarray}
where $\hat{H}_1 = (E\delta_{ij} + \gamma\eta_{ij})\x_j^i + \frac{\gamma}{2}\eta_{ij}(\x^{ij} + \x_{ij})$. Here we have assumed the condensate wavefunctions $\psi_i = \sqrt{N_i/V}$ are real, and $\delta_{ij}$ is the Kronecker delta. Thus, the parameter $E$ controls the kinetic term, and the dimensionless parameter $\gamma$ controls the hopping between the excitations and the condensate in the zero momentum. The $\hat{H}_2$ is chosen such that when $\tilde{g}=0$ the above driving process will bring the initial state to itself after two period of evolution. 

\begin{figure}
  \includegraphics[width = .45\textwidth]{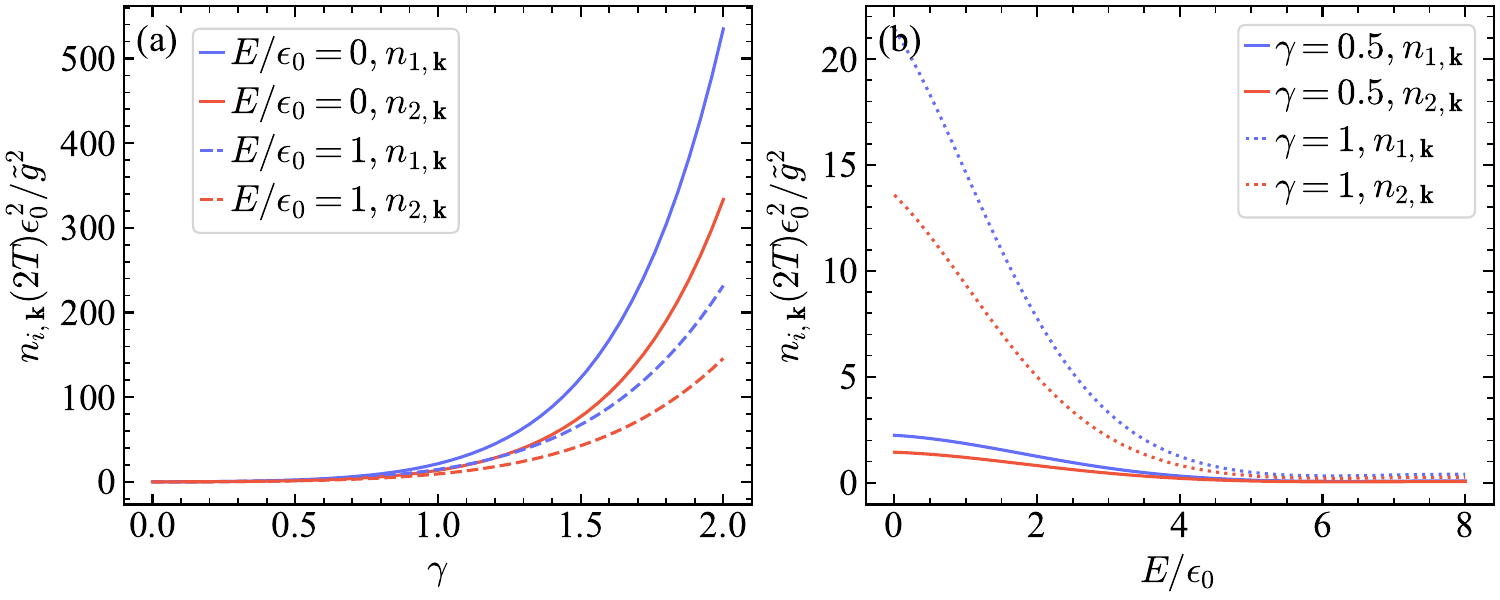}
  \caption{Particle numbers at the end of the $2T$ for each species of boson with (a) varying $\gamma$ and (b) varying $E$. For all these cases, the matrix $\eta = \left(\begin{array}{cc} 1/2 & 1/6 \\ 1/6 & 1/4 \end{array} \right)\epsilon_0$, where $\epsilon_0 = 2\eta_{11}$ serves as an energy unit.}
  \label{num_int}
\end{figure}

When $\tilde{g}\neq 0$, and $\tilde{g}$ is much smaller than the energy scale of $\hat{H}_1$, we can treat the interaction term perturbatively. By keeping the Dyson series to the first order, we have
\begin{eqnarray}
  \hat{U}_1' &=& e^{-i\hat{H}_1' t_1} \nn\\
  &=& e^{-i \hat{H}_1 t_1}\left[1 - i\tilde{g}\int_0^{t_1} \hat{V}(t) dt + O(U^2)\right],
\end{eqnarray}
where $\hat{V}(t) = e^{i\hat{H}_1 t} \hat{V} e^{-i\hat{H}_1 t}$. Since $\hat{H}_1$ is a linear combination of the $\mathfrak{sp}(4,R)$ Lie algebra, according to Ref.\cite{perelomov_generalized_1986, wang_quantum_2022}, the operator $\Psi_\bk$ evolves as
\begin{eqnarray}\label{psi_t}
  e^{i\hat{H_1}t}\Psi_\bk e^{-i\hat{H_1}t} = e^{-i\kappa\h_1 t}\Psi_\bk.
\end{eqnarray}
With this transformation, $\hat{V}(t)$ can be calculated. Then, the time evolution operator after two period of driving up to the first order in $\tilde{g}$ has the following form
\begin{eqnarray}
  \hat{U}(2T) &=& \hat{U}_2\hat{U}_1'\hat{U}_2\hat{U}_1' \nn\\
   &=& \left[1 - i\tilde{g}\int_0^{t_1}dt \hat{A} \hat{V}(t) \hat{A}^\dag \right. \nn\\
   && \left. - i\tilde{g}\int_0^{t_1} dt \hat{A}^2 \hat{V}(t) (\hat{A}^\dag)^2 + O(\tilde{g}^2)\right]\hat{A}^2 \nn\\
  & \equiv& [1 - i\tilde{g}\hat{\mathscr{V}} + O(\tilde{g}^2)]\hat{A}^2,
\end{eqnarray}
where we have defined $\hat{A} = e^{-i\hat{H}_2 t_2} e^{-i\hat{H}_1 t_1}$, and denoted the integral terms as $\hat{\mathscr{V}}$. Here, $\hat{A} \hat{V}(t) \hat{A}^\dag$ and $\hat{A}^2 \hat{V}(t) (\hat{A}^\dag)^2$ can also be calculated using the method in Eq.(\ref{psi_t}). As a result, the particle number at the end of the $2m$th period is given by
\begin{eqnarray}
  n_{i,\bk}(2mT) &=& \langle 0|[\hat{U}^\dag(2T)]^m \hat{n}_{i,\bk} [\hat{U}(2T)]^m |0\rangle \nn\\
   &=& m^2\tilde{g}^2\langle 0| \hat{\mathscr{V}}^\dag \hat{n}_{i,\bk} \hat{\mathscr{V}} |0\rangle + O(\tilde{g}^3)
\end{eqnarray}
where $\hat{n}_{i,\bk} = a_{i,\bk}^\dag a_{i,\bk}$. Here, we have used the relations $\hat{A}^2|0\rangle = -|0\rangle$ and $\hat{A}^2 \hat{n}_{i,\bk} (\hat{A}^\dag)^2 = \hat{n}_{i,\bk}$. Then, the expectation value part of above expression can be evaluated numerically. And we can conclude that the to the leading order, the particle numbers increase quadratically with interaction strengths $\tilde{g}$.

In Figure \ref{num_int}, we show the numerical results for the particle numbers $n_{i,\bk}(2T)$ at the end of the second driving period varying with the parameters $E$ and $\gamma$ with $\eta_{ij}$  fixed. From Figure \ref{num_int}(a), we can see that for different fixed values of $E$, the particle numbers increase exponentially with increasing $\gamma$. However, the particle numbers decrease as the parameter $E$ increases. This result can be understood as follows: $\gamma$ controls the hopping between the excitations and the condensate in zero momentum. When increasing $\gamma$ with fixed kinetic term parameter $E$, more particles will be excited. Since the echo-breaking interactions are between the excited particles. Thus, more excited particles will result stronger echo-breaking effect. On the other hand, when increasing $E$ with fixed $\gamma$, the ratio $\gamma/E$ gets smaller, thus, the particles exciting is suppressed. Hence, the echo-breaking effect is weaker in this case.

\section{Possible experiment implementation} \label{experiment}

\begin{figure}
  \includegraphics[width = .35\textwidth]{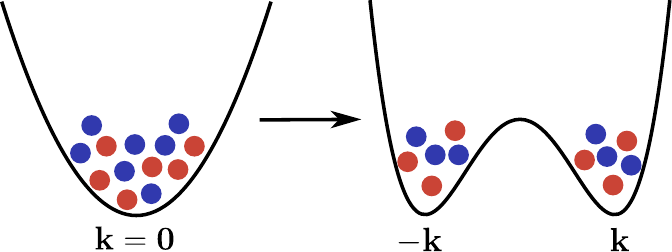}
  \caption{Schematics for the shaken optical lattice implementation of hopping terms between opposite momenta of two-component BEC Hamiltonian. Red and blue dots represent two species of bosons.}
  \label{shaken_lat}
\end{figure}
In previous sections, we explored a general driving protocol to facilitate quantum echo in two-component Bose-Einstein Condensates (BECs) and examined the influence of symmetry-breaking interactions on this phenomenon. In this section, we will propose a possible experiment implementation for this quantum echo. 

The model Eq.(\ref{ham}) involves the hopping between the bosonic modes in opposite momenta $\pm \bk$. The similar hopping also appears in the single-component BEC case and is proposed to be implemented in a double well structure in momentum space in Ref. \cite{lyu_geometrizing_2020}. This kind of double well structure can be implemented in various experimental setups such as shaken optical lattice, spin-orbit coupling and periodic driving \cite{lyu_geometrizing_2020}. For the two-component case, we propose using a shaken optical lattice \cite{parker_direct_2013} setup. The experimental procedure would start with preparing a two-component BEC, initially condensing all particles in the zero momentum state. Subsequently, altering the band structure to form a double-well would populate the particles in opposite momenta as shown in Figure \ref{shaken_lat}. This process corresponds to the hopping described in Equation (\ref{ham}) from opposite momenta.

In the second step of our driving protocol, the coupling term between opposite momenta disappear, however, the spin-orbit coupling terms like $a_{i,\bk}^\dag a_{j,\bk}$ with $i\neq j$ are always present. Importantly, the implementation of spin-orbit coupling is also feasible in cold atom experiments \cite{galitski_spin_2013}. Therefore, the driving protocol we propose can be effectively realized within the current frameworks of cold atom experimental setups.

\section{Conclusion}\label{conclusion}
In this paper, we have addressed the question of how quantum echo can be achieved in a two-component Bose-Einstein condensate system using the $Sp(4,R)$ group formalism. We have shown that under a periodic driving protocol consisting of two steps in each period, any $Sp(4,R)$ coherent state of two-component BEC can reverse to itself after two periods of driving. We have also developed a general method to construct the second-step Hamiltonian $\hat{H}_2$ from the first-step Hamiltonian $\hat{H}_1$ based on the property of real symplectic matrices and applied our method to two examples, one with time-independent $\hat{H}_1$ and one with time-dependent $\hat{H}_1$, then verified our theoretical prediction numerically. Furthermore, we have investigated how the interaction effect breaks the quantum echo process and found that the particle number deviation from the perfect revival is influenced by the ratio between the strengths of kinetic terms and interaction in the first-step Hamiltonian $\hat{H}_1$.

Our work extends the previous studies on quantum echo in single-component BEC and reveals new features of two-component BEC dynamics. Our work also showcases the usefulness and elegance of the $Sp(4,R)$ group formalism for studying quantum systems with real symplectic structures. Since our proof on the occurrence of quantum echo in two-component BEC system is solely based on the symplectic properties of the time evolution matrix, our result can also have potential use in spinor BEC and $N$-component boson system. Meanwhile, given the presence of real symplectic matrices in various areas of physical study \cite{weedbrook_gaussian_2012, simon_pereshorodecki_2000, simon_gaussian_1988,colas_fourmode_2022, alhassid_group_1983, deenen_boson_1985, enayati_antide_2023, penna_twospecies_2017, richaud_quantum_2017, charalambous_control_2020}, we hope that our work will stimulate further research on this topic and inspire new applications of quantum echo in ultracold atom systems and beyond.

\begin{acknowledgements}
CYW thanks Yan He for valuable discussions and suggestions on this manuscript. CYW is supported by the Shuimu Tsinghua scholar program at Tsinghua University.

\end{acknowledgements}

\appendix
\begin{widetext}

\section{The real symplectic group $Sp(2n, R)$} \label{app:sp}
In this appendix, we will give a brief introduction to the real symmetric group $Sp(2n, R)$, where $n$ is positive integer, and $R$ denotes the field of real numbers. When $n=2$, we have the group $Sp(4,R)$. More details can be found in \cite{hasebe_sp_2020}. 

The real symplectic group $Sp(2n, R)$ is the set of $2n\times 2n$ real matrices preserving a non-singular skew symmetric matrix $\Omega$, i.e. for a real matrix $M\in Sp(2n,R)$, it satisfies 
\begin{eqnarray}\label{symp}
  M^T\Omega M = \Omega,
\end{eqnarray}
where $M^T$ is the transpose of $M$. Here, the matrix $\Omega$ is defined as 
\begin{eqnarray}
  \Omega = \left(\begin{array}{cc} 
    0 & I_n \\
   -I_n & 0
  \end{array}\right),
\end{eqnarray}
where $I_n$ is the $n\times n$ identity matrix. Then, one can diagonalize the matrix $\Omega$ by a unitary transformation
\begin{eqnarray}\label{diag_trans}
  U \Omega U^\dag = i\kappa,
\end{eqnarray}
where $\kappa$ is a diagonal matrix
\begin{eqnarray}
  \kappa = \left(\begin{array}{cc} I_n & 0 \\ 0 & -I_n \end{array} \right),
\end{eqnarray}
and the unitary transformation $U$ is given by
\begin{eqnarray}
  U = \frac{1}{\sqrt{2}}\left(\begin{array}{cc} R & -iR \\ R & iR 
  \end{array} \right),
\end{eqnarray}
where $R$ is $n\times n$ matrix 
\begin{eqnarray}
  R = \left(\begin{array}{ccccc} 
    0 & 0 & \cdots & 0 & 1 \\
    0 & 0 & \cdots & 1 & 0 \\
    \vdots & \vdots & 1 & \vdots & \vdots \\
    0 & 1 & \cdots & 0 & 0 \\
    1 & 0 & \cdots & 0 & 0 \\
  \end{array} \right)
\end{eqnarray}
Then, by substituting Eq.(\ref{diag_trans}) to Eq.(\ref{symp}), we have
\begin{eqnarray} \label{g_tilde}
  (U M U^\dag)^\dag \kappa (U M U^\dag) = \kappa,
\end{eqnarray}
where we have used the fact that $M$ is a real matrix. Thus, by defining $\mathcal{M}$ as 
\begin{eqnarray}
  \mathcal{M} = U M U^\dag,
\end{eqnarray}
we have another representation of the $Sp(2n,R)$ group. And $\mathcal{M}$ satisfies the constraint 
\begin{eqnarray}\label{constraint}
  \mathcal{M}^\dag \kappa \mathcal{M} = \kappa.
\end{eqnarray} 
By substituting Eq.(\ref{diag_trans}) to Eq.(\ref{g_tilde}), we can see that the matrix $\mathcal{M}$ has the form 
\begin{eqnarray}
  \mathcal{M} = \left(\begin{array}{cc} \cua & \cva \\ \cva^* & \cua^*,
  \end{array}\right)
\end{eqnarray}
where $\cua$ and $\cva$ are $n\times n$ matrices. And the constraint Eq.(\ref{constraint}) becomes 
\begin{eqnarray}
  \cua\cua^\dag - \cva\cva^\dag = I_n, \ \ \cua\cva^T = \cva^T\cua.
\end{eqnarray}

Note that in this representation, the matrix $\mathcal{M}$ is no longer real matrix in general. However, this representation is suitable for describing the quantum dynamics of two-component BEC.

\section{The explicit expression of $\cua_0$}\label{app1}
In this appendix, we will give the explicit expression of $\cua_0$ of the matrix $U_0 = e^{\frac{i}{2}(\eta_{ij} Y^{ij} + \eta_{ij}^* Y_{ij})}$. The explicit form of the matrix generators $\{Y_{ij}, Y^{ij}, Y_l^k\}$ is \cite{hasebe_sp_2020}
\begin{eqnarray}
  &&(\kappa Y_{ij})_{ab} = \delta_{a,2+i}\delta_{b,j} +\delta_{a,2+j}\delta_{b,i}, \nn\\
  &&(\kappa Y^{ij})_{ab}= \delta_{a,i}\delta_{b,2+j}+\delta_{a,j}\delta_{b,2+i}, \nn\\
  &&(\kappa Y_l^k)_{ab} = \delta_{a,k}\delta_{b,l}+\delta_{2+k,b}\delta_{2+l,a},
\end{eqnarray}
where $\kappa = \mathrm{diag}(1,1,-1,-1)$. Then, we define 
\begin{eqnarray}
  \mathcal{A} = \frac{1}{2}\eta_{ij} Y^{ij} + \frac{1}{2}\eta_{ij}^* Y_{ij} = \left(\begin{array}{cc} 0 & \eta \\ -\eta^* & 0 \end{array}\right).
\end{eqnarray}
According to Ref.\cite{wang_quantum_2022}, its eigenvalues are of the form $\{\lambda_+,\lambda_-,-\lambda_+,-\lambda_-\}$, where 
\begin{eqnarray}\label{eigen}
  \lambda_\pm = \frac{1}{2}\sqrt{\mathrm{Tr}(\mathcal{A}^2) \pm \sqrt{\mathrm{Tr}(\mathcal{A}^2)^2 - 16\det(\mathcal{A})}}.
\end{eqnarray}
It can be calculated directly by Mathematica that $\cua_0$ is given as 
\begin{eqnarray}
  \cua_{0,11} &=& \frac{|\eta_{11}|^2 - |\eta_{22}|^2}{2(\lambda_+^2-\lambda_-^2)}[\cos(\lambda_-) - \cos(\lambda_+)] + \frac{1}{2}[\cos(\lambda_-) + \cos(\lambda_+)], \nn\\
  \cua_{0,12} &=& \frac{\eta_{11}\eta_{12}^* + \eta_{22}^*\eta_{12}}{\lambda_+^2-\lambda_-^2}[\cos(\lambda_-) - \cos(\lambda_+)], \nn\\
  \cua_{0,21} &=& \frac{\eta_{11}^*\eta_{12} + \eta_{22}\eta_{12}^*}{\lambda_+^2-\lambda_-^2}[\cos(\lambda_-) - \cos(\lambda_+)], \nn\\
  \cua_{0,22} &=& -\frac{|\eta_{11}|^2 - |\eta_{22}|^2}{2(\lambda_+^2-\lambda_-^2)}[\cos(\lambda_-) - \cos(\lambda_+)] + \frac{1}{2}[\cos(\lambda_-) + \cos(\lambda_+)].
\end{eqnarray}
If $\cua_0$ is Hermitian, $\cua_{0,11}$ and $\cua_{0,22}$ should be real, meanwhile $\cua_{0,12} = \cua_{0,21}^*$. These conditions lead to that $\cos(\lambda_+), \cos(\lambda_-)$ and $\lambda_+^2 - \lambda_-^2$ should be real numbers. On the other hand, from Eq.(\ref{eigen}), we have
\begin{eqnarray}
  \lambda_+^2 - \lambda_-^2 &=& \frac{1}{2}\sqrt{\mathrm{Tr}(\mathcal{A}^2)^2 - 16\det(\mathcal{A})} \nn\\
   &=& \frac{1}{2}\sqrt{4(|\eta_{11}|^2 + |\eta_{22}|^2 + 2|\eta_{12}|^2)^2 - 16|\eta_{12}^2 - \eta_{11}\eta_{22}|^2} \nn\\
   &=& \sqrt{(\eta_{11}|^2 + |\eta_{22}|^2 + 2|\eta_{12}|^2 + 2|\eta_{12}^2 - \eta_{11}\eta_{22}|)(\eta_{11}|^2 + |\eta_{22}|^2 + 2|\eta_{12}|^2 - 2|\eta_{12}^2 - \eta_{11}\eta_{22}|)} \nn\\
   &\geq& \sqrt{(\eta_{11}|^2 + |\eta_{22}|^2 + 2|\eta_{12}|^2 + 2|\eta_{12}^2 - \eta_{11}\eta_{22}|)(\eta_{11}|^2 + |\eta_{22}|^2 + 2|\eta_{12}|^2 - 2|\eta_{12}|^2 - 2|\eta_{11}||\eta_{22}|)} \nn\\
   &=& \sqrt{(\eta_{11}|^2 + |\eta_{22}|^2 + 2|\eta_{12}|^2 + 2|\eta_{12}^2 - \eta_{11}\eta_{22}|)(\eta_{11}|^2 - |\eta_{22}|^2)^2 } \nn\\
   &\geq& 0.
\end{eqnarray}
Hence, $\lambda_+^2 - \lambda_-^2$ is real. Meanwhile, since $\mathrm{Tr}(\mathcal{A}^2) = -2(|\eta_{11}|^2 + |\eta_{22}|^2 + 2|\eta_{12}|^2)$, according to Eq.(\ref{eigen}), both $\lambda_+$ and $\lambda_-$ are imaginary. Thus, $\cos(\lambda_+)$ and $\cos(\lambda_-)$ are real. Hence, the conditions of $\mathcal{U}_0$ being Hermitian are all satisfied. As a result, we have proven that $\cua_0$ is a $2\times 2$ Hermitian matrix.

\end{widetext}

\bibliography{ref}

\end{document}